\RequirePackage{fix-cm}
\documentclass[fleqn,usenatbib]{mnras}
\usepackage{newtxtext,newtxmath,graphicx,amsmath,hyperref,xcolor,array,multirow}
\usepackage[T1]{fontenc}

\DeclareRobustCommand{\VAN}[3]{#2}
\let\VANthebibliography\thebibliography
\def\thebibliography{\DeclareRobustCommand{\VAN}[3]{##3}\VANthebibliography}

\newcolumntype{?}{!{\vrule width 3pt}}

\title[Nonparametric curvature with DESI DR2]{Stress-testing the spatially flat Universe: nonparametric spatial curvature determination from DESI DR2}

\author[H. C. Turner et al.]{Hannah C. Turner,$^{1}$\thanks{E-mail: h.c.turner@bham.ac.uk} Suhail Dhawan,$^{1}$ Sunny Vagnozzi,$^{2,3}$ Alice Townsend$^{1}$ and Peter Massey$^{1}$ \\
$^{1}$School of Physics and Astronomy, University of Birmingham, Edgbaston Park Road, Birmingham B15 2TT, United Kingdom\\
$^{2}$Department of Physics, University of Trento, Via Sommarive 14, 38123 Povo (TN), Italy\\
$^{3}$Trento Institute for Fundamental Physics and Applications (TIFPA)-INFN, Via Sommarive 14, 38123 Povo (TN), Italy}

\date{Accepted XXX. Received YYY; in original form ZZZ}

\pubyear{\the\year{}}

\begin{document}
\label{firstpage}
\pagerange{\pageref{firstpage}--\pageref{lastpage}}
\maketitle

\begin{abstract}
\noindent We present a nonparametric determination of spatial curvature using only late-time geometrical probes. Under a Gaussian Process smoothness prior, we jointly constrain the dimensionless expansion history and $\Omega_K$, combining three Type Ia Supernovae (SNeIa) samples with either \textit{DESI} DR2 line-of-sight baryon acoustic oscillations (BAO) or cosmic chronometers (CC) data. From our baseline SNeIa+BAO analyses we find $\Omega_K=0.219^{+0.110}_{-0.111}$, $\Omega_K=0.330^{+0.098}_{-0.150}$ and  $\Omega_K=0.260^{+0.124}_{-0.127}$ for the \textit{PantheonPlus}, \textit{DES Dovekie} and \textit{Union3} SNeIa samples respectively. These nonparametric results mildly favour an open Universe, supporting recent hints in this direction, and improve on the sensitivity of a previous application of the method by up to a factor of $\sim 2.5$. We find consistent constraints combining SNeIa and CC, whereas restricting SNeIa data to the redshift range covered by BAO somewhat weakens the constraints and shifts the \textit{PantheonPlus} central value to $\Omega_K<0$. Our results depend only on late-time relative distances and expansion rates, making them insensitive to the absolute distance scale calibration, while relying neither on Cosmic Microwave Background data nor on a parametric dark energy model.
\end{abstract}

\begin{keywords}
cosmological parameters -- cosmology:observations -- distance scale
\end{keywords}

\section{Introduction}
\label{sec:introduction} 

One of the most fundamental questions in cosmology concerns the local geometry of the observable Universe. Assuming homogeneity and isotropy on large scales, this amounts to measuring the spatial curvature parameter $\Omega_K$. Aside from determining the intrinsic geometry of spatial hypersurfaces, $\Omega_K$ carries valuable information about the primordial Universe, particularly the dynamics of inflation~\citep{Qiu:2026npi}, while also affecting the relation between late-time expansion history and distances. The importance of high-fidelity measurements of $\Omega_K$ which are as model-independent as possible cannot therefore be overstated, yet most available constraints depend on the assumed form of the late-time expansion history.~\footnote{See for instance~\cite{Vagnozzi:2017ovm,Wang:2019yob,Efstathiou:2020wem,Chudaykin:2020ghx,Benisty:2020otr,Vagnozzi:2020rcz,Vagnozzi:2020dfn,DiValentino:2020kpf,Yang:2021hxg,Cao:2021ldv,Arjona:2021hmg,Gonzalez:2021ojp,Dinda:2021ffa,Zuckerman:2021kgm,Bargiacchi:2021hdp,Akarsu:2021max,Glanville:2022xes,Bel:2022iuf,Wu:2022fmr,Yang:2022kho,deCruzPerez:2022hfr,Stevens:2022evv,Favale:2023lnp,Qi:2023oxv,Giare:2023ejv,Park:2024vrw,Forconi:2025zzu,Specogna:2025ufe,Favale:2025mgk,Comini:2026nsj} for examples of recent $\Omega_K$ constraints.}

This question is especially timely now, as evidence for a dynamical dark energy (DE) component is mounting in light of recent Baryon Acoustic Oscillation (BAO) and Type Ia Supernovae (SNeIa) data. The same data appear to mildly favour negative spatial curvature, i.e.\ a spatially open Universe with $\Omega_K>0$~\citep[see for instance][]{Mukherjee:2024ryz,Jiang:2024xnu,Wu:2024faw,Li:2024yrg,Liu:2024yib,Chen:2025mlf,Chaudhary:2025uzr,Gong:2025hoy,Capozziello:2025lor,Chudaykin:2025lww,Yadav:2025wbc,Wang:2025hvo,Wu:2026qog,Giare:2026oti,Chudaykin:2026amr}, opposite in sign to the positive spatial curvature (spatially closed Universe, $\Omega_K<0$) previously suggested by \textit{Planck} 2018 Cosmic Microwave Background (CMB) data on their own~\citep{Handley:2019tkm,DiValentino:2019qzk,DiValentino:2020hov}. To add to the confusion, the apparent preference for $\Omega_K>0$ is tied to a number of other mild tensions concerning the matter density parameter $\Omega_m$~\citep{Sakr:2023hrl,Baryakhtar:2024rky,Pedrotti:2024kpn,Colgain:2024mtg,Weiner:2026sfm,Shlivko:2026jxa,Schoneberg:2026buf}, the neutrino mass sum $\sum m_{\nu}$~\citep[][]{Naredo-Tuero:2024sgf,Jiang:2024viw,Lynch:2025ine,Graham:2025dqn,Kibris:2026cqq}, and the optical depth to reionization $\tau$~\citep[][]{Sailer:2025lxj,Jhaveri:2025neg}; in each case, the statistical significance depends to some extent on the assumed expansion history. Overall, the current situation has generated some confusion regarding the observational status of $\Omega_K$, and urgently calls for a determination thereof which is as purely geometrical and model-independent as possible.

In the earlier work of~\cite{Dhawan:2021mel}, some of us introduced a framework to infer purely geometrical, nonparametric constraints on $\Omega_K$, requiring only three ingredients: \textit{relative} distances from unanchored SNeIa, \textit{relative} expansion rate information provided by either cosmic chronometers (CC) or line-of-sight BAO measurements, and a Gaussian Process (GP) smoothness prior on the expansion history. In short, the method can be understood as implicitly comparing relative distances measured by SNeIa with the distance-redshift relation obtained by integrating the GP-smoothed relative expansion history, as the relation between the two is controlled by $\Omega_K$. The method was applied to then-current \textit{Pantheon} SNeIa and CC data, yielding $\Omega_K=-0.03 \pm 0.26$. Available radial BAO data were limited to three measurements over a narrow redshift range, so their potential was studied only through a Stage IV forecast.

Our \textit{Letter} is motivated by the growing confusion outlined above, the need to urgently clarify the observational status of $\Omega_K$, and especially the significantly wider data landscape compared to 2021. This includes the first Stage IV BAO data from \textit{DESI}, as well as three distinct SNeIa samples: the \textit{PantheonPlus}, Dark Energy Survey Year 5 (\textit{DESY5}), and \textit{Union3} ones. We exploit these to carry out the previous forecast on real data, using the three SNeIa datasets to assess the robustness of our results against the choice of sample, while performing a redshift-overlap test to study our sensitivity to redshift coverage, and updating the earlier SNeIa+CC inference in light of a much better handle on the systematics and covariance of CC measurements. In short, we find a factor of up to $2.5$ improvement in sensitivity compared to the earlier results of~\cite{Dhawan:2021mel}, with all three baseline combinations favoring $\Omega_K>0$. Importantly, since only relative distances and expansion rates enter in our methodology, the resulting constraints on $\Omega_K$ are insensitive to the calibration of our probes. This is particularly valuable in the context of the Hubble tension, as disagreements on the absolute distance scale do not propagate into our inference.

\section{Datasets and methodology}
\label{sec:datasets}

We make use of the following state-of-the-art observations:
\begin{itemize}
\item SNeIa distance moduli measurements from the \textit{PantheonPlus} sample of spectroscopically classified SNeIa~\citep{Brout:2022vxf}. As we are interested in using SNeIa as relative distance indicators, we exclude the low-redshift calibrator sample, and only use measurements in the redshift range $0.01<z<2.26$.
\item SNeIa distance moduli measurements from the \textit{DESY5} sample of photometrically classified SNeIa, recalibrated through the \textit{Dovekie} cross-calibration program~\citep{DES:2025sig}, in the redshift range $0.025<z<1.13$. We refer to these measurements as \textit{DES Dovekie}.
\item SNeIa distance moduli measurements from the \textit{Union3} sample of spectroscopically classified SNeIa, binned in 22 bins in the redshift range $0.05<z<2.26$~\citep{Rubin:2023jdq}.
\item Line-of-sight BAO measurements of $D_H/r_d$ from the \textit{DESI} DR2 release~\citep{DESI:2025zgx}. Since we only use $D_H/r_d$ measurements, we exclude the $z=0.295$ \textit{BGS} sample, for which only a volume-averaged distance measurement is available. Our dataset therefore covers 6 redshift bins in the range $0.510<z<2.330$.
\item 15 \textit{CC} measurements of $H(z)$ in the range $0.179<z<1.965$ from the relative ages of massive, early-type, passively-evolving galaxies~\citep{Moresco:2012by,Moresco:2015cya,Moresco:2016mzx}. We only use the subset of measurements for which the full covariance matrix is available, including non-diagonal terms and contributions from systematics following~\citet{Moresco:2018xdr,Moresco:2020fbm}.
\end{itemize}
The three SNeIa samples form the backbone of our analysis. However, because we infer $\Omega_K$ via the consistency between two independently measured functions of redshift, we need to ensure that our result is not driven by a particular SNeIa sample, expansion rate probe, or mismatch in their redshift coverage. To address these questions, for each sample we consider three dataset combinations: SNeIa+BAO over the full redshift range, SNeIa+BAO restricting SNeIa to the redshift range covered by BAO (i.e.\ considering only $z>0.51$ SNeIa), and SNeIa+CC. We therefore study a total of nine dataset combinations, which allow us to assess the robustness of our results against choice of SNeIa sample, relative expansion rate probe, and redshift coverage. The redshift overlap test is important to examine the extent to which our $\Omega_K$ inference is sensitive to the low-redshift SNeIa anchoring of the relative distance relation, the SNeIa-driven GP reconstruction of $H(z)$ below the first BAO point, as well as possible inconsistencies between the low- and high-$z$ parts of each SNeIa sample.

We recall that SNeIa apparent magnitudes $m$ are related to the luminosity distance $d_L(z)$ as follows:
\begin{equation}
m(z)=5\log_{10} \left [ \frac{d_L(z)}{{\text{Mpc}}} \right ] +M+25\,,
\label{eq:mu}
\end{equation}
with $M$ the SNeIa absolute magnitude, and $d_L(z)$ given by:
\begin{equation}
d_L(z)=\frac{c(1+z)}{H_0\sqrt{\vert \Omega_K \vert}}\operatorname{sinn} \left \{ \sqrt{\vert \Omega_K \vert}\int_0^z\frac{dz'}{E(z')} \right \} \,,
\label{eq:dl}
\end{equation}
where $H_0$ is the Hubble constant, $E(z) \equiv H(z)/H_0$ is the dimensionless expansion rate, and $\text{sinn}(x)=[\sin(x),\sinh(x),x]$ for $[\Omega_K<0,\Omega_K>0,\Omega_K=0]$.

Conceptually, the methodology of~\citet{Dhawan:2021mel} first nonparametrically reconstructs $E(z)$ from CC and radial BAO data. With $E(z)$ determined, the integral in Eq.~(\ref{eq:dl}) is calculated numerically, leaving $\Omega_K$ as the only remaining unknown to be determined by comparing the theoretical prediction for $\mu$ in Eq.~(\ref{eq:mu}) to the measured distance moduli. Of course, this two-step procedure is not what we do in practice. Rather, we impose a GP smoothness prior on $H(z)$, and jointly sample the posterior distribution of $\Omega_K$, $M$, $H(z)$, and the GP hyperparameters. Following~\citet{Dhawan:2021mel}, we discretize $H(z)$ at a number of nodes (whose redshifts are free parameters we marginalize over), so the GP prior on $H(z)$ turns into a multivariate Gaussian prior on $H(z)$ evaluated at the nodes. We adopt a squared-exponential kernel, and marginalize over the amplitude and length scale of the prior covariance. We sample the joint posterior of $\Omega_K$, $M$, $H(z)$ evaluated at the nodes (and conditioned on the GP hyperparameters) using Hamiltonian Monte Carlo sampling, implemented in \texttt{PyStan}~\citep{pystan}, and setting uniform priors on $\Omega_K$ and $M$. As our methodology is identical to the one of~\citet{Dhawan:2021mel}, our summary is deliberately brief, and we refer the reader to~\cite{Dhawan:2021mel} for a complete discussion thereof.

We stress that our methodology is insensitive to the absolute expansion rate calibration. This is true even when CC data is included: despite their measuring $H(z)$, they only influence our determination of $\Omega_K$ through $E(z)$. Likewise, with BAO data, we can safely fix the sound horizon to the $\Lambda$CDM value $r_d=147\,{\text{Mpc}}$. As in~\cite{Dhawan:2021mel}, we explicitly verify our insensitivity to the overall calibration of SNeIa, CC, or BAO, by rescaling the datasets by arbitrary constants (amounting to changing $M$, the normalization of $H(z)$, or $r_d$ respectively), finding identical constraints on $\Omega_K$. This sensitivity to the ``shape'' of the expansion history $E(z)$, rather than the overall calibration, is very important given the Hubble tension~\citep[][]{DiValentino:2021izs,CosmoVerseNetwork:2025alb,Cai:2026swf}, as uncertainties on the absolute distance scale do not propagate into our inference of $\Omega_K$.~\footnote{The importance of shape constraints in relation to the Hubble tension has recently been discussed in~\citet{Zhou:2025kws,Pedrotti:2025ccw,Bansal:2026axl,Zhou:2026iar,Sabogal:2026ipu}.}

Other recent studies have used GPs to infer $\Omega_K$ nonparametrically from late-time data alone~\citep[see][]{Yang:2020bpv,Wu:2022fmr,Favale:2023lnp,Qi:2023oxv,Gong:2024ugr,Liu:2024yib,Dias:2024tpf,Gao:2025ozb,Gong:2025hoy,Favale:2025mgk}. While sharing the same overall (geometrical) motivation, our work differs substantially from each of these in either or both the choice of datasets and statistical implementation. Many of these studies relied on pre-\textit{DESI} or \textit{DESI} DR1 data, neglected the CC covariance, were sensitive to the absolute calibration, or optimized and fixed the GP hyperparameters rather than marginalizing over them. While \cite{Favale:2025mgk} is closest to our work in their marginalizing over the GP hyperparameters and using the full CC covariance, they use CC as an absolute expansion rate anchor, and do not perform a comparative analysis across SNeIa samples. To the best of our knowledge, ours is the first nonparametric $\Omega_K$ determination which is insensitive to the absolute calibration of all probes, combines all the latest (as of 2026) available data, while explicitly marginalizing over the GP hyperparameters, and testing the impact of redshift coverage.

\section{Results}
\label{sec:results}

\begin{figure}
\centering
\includegraphics[width=0.97\linewidth]{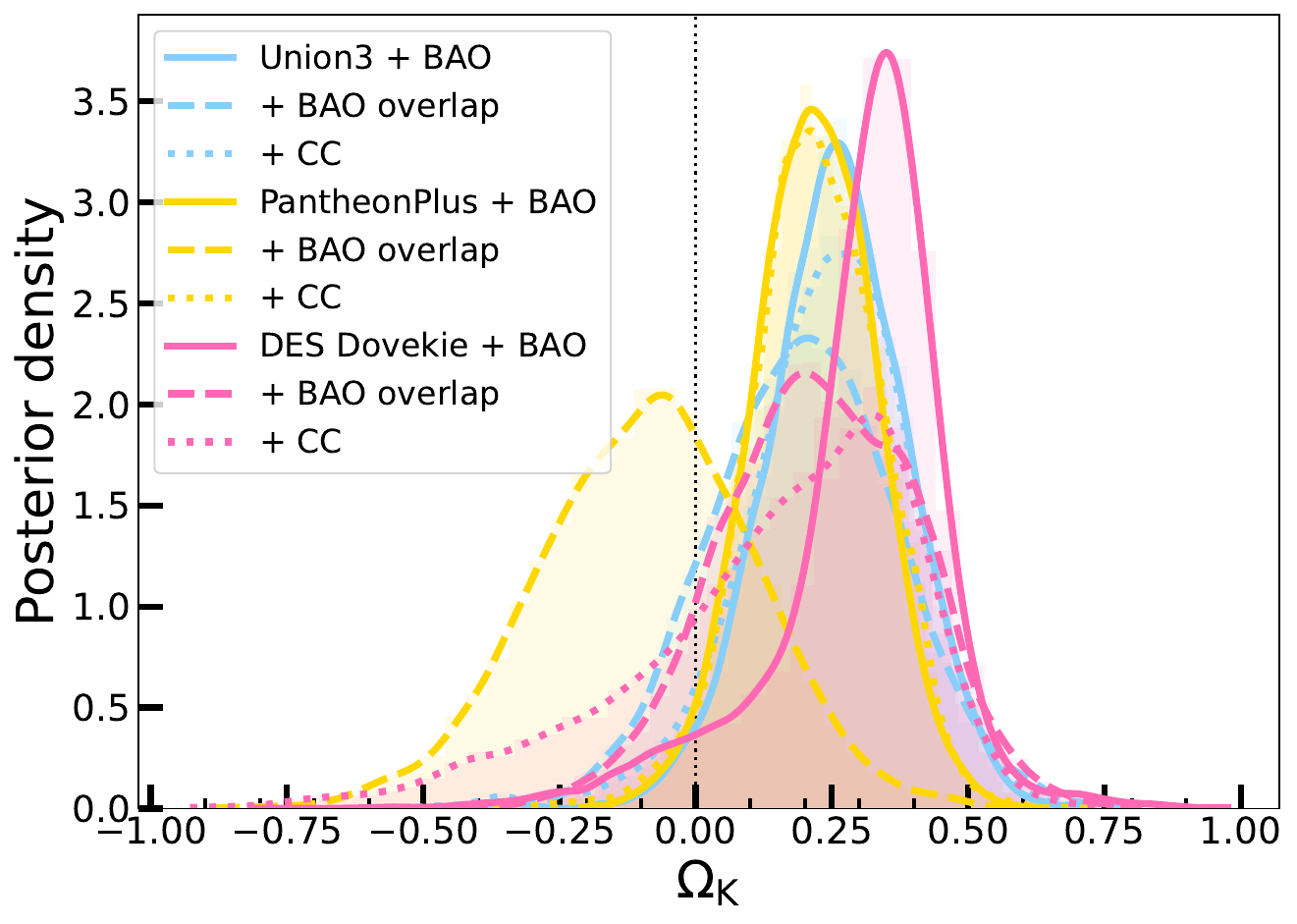}
\caption{1D marginalized posterior distributions for the spatial curvature parameter $\Omega_K$, obtained in light of the nine dataset combinations considered in this work. For each of the three SNeIa samples (\textit{Union3} in blue, \textit{PantheonPlus} in yellow, and \textit{DES Dovekie} in pink), we consider three dataset combinations: SNeIa+BAO over the full redshift range (solid curves), SNeIa+BAO restricting SNeIa to the redshift range covered by BAO (i.e.\ considering only $z>0.51$ SNeIa, dashed curves), and SNeIa+CC (dotted curves). The vertical dotted black line indicates the spatially flat case $\Omega_K=0$.}
\label{fig:posteriors}
\end{figure}

Our constraints on $\Omega_K$ in light of the nine dataset combinations analysed are presented in Tab.~\ref{tab:results}. Marginalized posterior distributions for $\Omega_K$ obtained in light of the nine dataset combinations considered are shown in Fig.~\ref{fig:posteriors}, where the SNeIa compilations are distinguished by colors (blue, yellow, and pink for \textit{Union3}, \textit{PantheonPlus}, and \textit{DES Dovekie} respectively), whereas the external expansion rate probes are instead distinguished by line style (solid, dashed, and dotted for full BAO sample, $z>0.51$ BAO, and CC respectively).

\begin{table}
\centering
\footnotesize
\resizebox{0.97\linewidth}{!}{
\begin{tabular}{|c|c|c?c|}
\hline
\textbf{SNeIa sample} & \textbf{Expansion rate probe} & \textbf{SNeIa cut} & $\boldsymbol{\Omega_K}$ \\
\hline\hline
\multirow{3}{*}{\textit{PantheonPlus}} & BAO & Full sample & $0.219^{+0.110}_{-0.111}$ \\
\cline{2-4} & BAO & $z>0.51$ only & $-0.085^{+0.191}_{-0.202}$ \\
\cline{2-4} & CC & Full sample & $0.219^{+0.120}_{-0.116}$ \\
\hline\hline
\multirow{3}{*}{\textit{DES Dovekie}} & BAO & Full sample & $0.330^{+0.098}_{-0.150}$ \\
\cline{2-4} & BAO & $z>0.51$ only & $0.226^{+0.181}_{-0.180}$ \\
\cline{2-4} & CC & Full sample & $0.197^{+0.183}_{-0.304}$ \\
\hline\hline
\multirow{3}{*}{\textit{Union3}} & BAO & Full sample & $0.260^{+0.124}_{-0.127}$ \\
\cline{2-4} & BAO & $z>0.51$ only & $0.200^{+0.166}_{-0.170}$ \\
\cline{2-4} & CC & Full sample & $0.250^{+0.134}_{-0.151}$ \\
\hline
\end{tabular}}
\caption{$68\%$ credible intervals on the spatial curvature parameter obtained in light of the nine dataset combinations considered in this work. For each of the three SNeIa samples, we consider three dataset combinations: SNeIa+BAO over the full redshift range, SNeIa+BAO restricting SNeIa to the redshift range covered by BAO (i.e.\ considering only $z>0.51$ SNeIa), and SNeIa+CC.}
\label{tab:results}
\end{table}

We begin with the SNeIa+BAO combinations over the full redshift range, shown by the solid curves in Fig.~\ref{fig:posteriors}, which constitute our baseline dataset combinations and provide our most constraining results. We infer $\Omega_K=0.219^{+0.110}_{-0.111}$, $\Omega_K=0.330^{+0.098}_{-0.150}$, and $\Omega_K=0.260^{+0.124}_{-0.127}$ when using the \textit{PantheonPlus}, \textit{DES Dovekie}, and \textit{Union3} SNeIa samples respectively. The three values are mutually consistent well within $1\sigma$ (see also the three solid posteriors in Fig.~\ref{fig:posteriors}), and all mildly favour an open Universe, with the spatially flat $\Omega_K=0$ case lying $\sim2\sigma$ away from the respective central values. The consistency across the three determinations indicates that the mild preference for $\Omega_K>0$ is very robust against the choice of SNeIa sample. However, since all three combinations contain the same BAO sample, the agreement should not be interpreted as three independent indications for $\Omega_K>0$. Compared to the earlier \textit{Pantheon}+CC constraint of $\Omega_K=-0.03 \pm 0.26$~\citep{Dhawan:2021mel}, our baseline results show an improvement in sensitivity to spatial curvature by a factor of approximately $2$--$2.5$.

Restricting the SNeIa samples to $z>0.51$, i.e.\ to the redshift range covered by radial BAO measurements substantially weakens our constraints on $\Omega_K$ as expected. We obtain $\Omega_K=-0.085^{+0.191}_{-0.202}$, $0.226^{+0.181}_{-0.180}$, and $0.200^{+0.166}_{-0.170}$ from the \textit{PantheonPlus}, \textit{DES Dovekie}, and \textit{Union3} samples respectively. The \textit{DES Dovekie} and \textit{Union3} central values remain positive and close to their baseline results. On the other hand, the \textit{PantheonPlus} central value becomes negative, $\Omega_K=-0.085^{+0.191}_{-0.202}$. There is no evidence for inconsistency across all three samples, but in all cases the preference for an open Universe weakens after the cut, especially for \textit{PantheonPlus}, indicating that there is some sensitivity to low-redshift SNeIa coverage. We stress that, even for $z>0.51$ SNeIa, the predicted luminosity distance still involves the integral of $1/E(z)$ from $z=0$, so this overlap test does not entirely isolate our curvature inference within the BAO range.

Replacing the radial BAO measurements with CC data provides a complementary test using an independent late-time background probe. In this case we find $\Omega_K=0.219^{+0.120}_{-0.116}$, $0.197^{+0.183}_{-0.304}$, and $0.250^{+0.134}_{-0.151}$ for the \textit{PantheonPlus}, \textit{DES Dovekie}, and \textit{Union3} samples respectively. We see that all three determinations are consistent with the baseline results discussed earlier, and therefore confirm the hint for $\Omega_K>0$. The \textit{PantheonPlus} and \textit{Union3} results retain a mild preference for an open Universe, while the broader and more asymmetric \textit{DES Dovekie}+CC posterior shows a weaker preference, with $\Omega_K=0$ lying within the $68\%$ credible interval. Nevertheless, the overall agreement between the BAO- and CC-based determinations indicates that the preference for $\Omega_K>0$ in the full SNeIa samples is \textit{not} specific to the choice of $H(z)$ probe.

Taken together, these results paint a broadly consistent picture. The hints for $\Omega_K>0$ persist across SNeIa samples and expansion rate probes, but explicitly depend on the available redshift lever arm. The tightest constraints come, unsurprisingly, from the baseline full SNeIa+BAO combinations, which are compatible with the CC-based determination and all favour $\Omega_K>0$. The preference weakens when restricting to $z>0.51$ SNeIa, with the \textit{PantheonPlus} central value changing sign. This overall consistency suggests that the indications for $\Omega_K>0$ are not driven entirely by the choice of SNeIa sample or $H(z)$ probe, although the \textit{PantheonPlus} redshift overlap results indicate some degree of sensitivity to redshift coverage.

\section{Discussion}
\label{sec:discussion}

Our results provide overall consistent, albeit not conclusive, hints for a spatially open Universe from late-time geometrical data, with the preference for $\Omega_K>0$ reaching in some cases the $\approx 3\sigma$ level. These hints are consistent across the three SNeIa samples and broadly compatible with the CC determinations, although the preference weakens when removing low-$z$ SNeIa. Our results therefore add to a growing body of post-\textit{DESI} literature presenting similar hints. These results are based only on late-universe distance measurements, and therefore have the added advantage of being independent of the CMB. All the cases we studied marginally prefer an open universe, except when restricting the \textit{PantheonPlus} compilation to the redshift range covered by BAO.

It is instructive to compare our findings to those of recent related studies, focusing on the post-\textit{DESI} literature. The most direct comparison is to works which infer $\Omega_K$ while reconstructing the late-time expansion history in a nonparametric or quasi-model-independent way (albeit with differences in methodology, as discussed in Sec.~\ref{sec:datasets}). Several such analyses have found hints for $\Omega_K>0$, consistent with ours. For instance, \cite{Jiang:2024xnu} find central values for $\Omega_K$ ranging between $0.067$ and $0.135$, with uncertainties of order $0.09$: these are slightly smaller than ours despite the use of \textit{DESI} DR1 BAO data, as a result of the slightly different GP methodology, which allowed for the use of transverse and volume-averaged BAO data as well. Using a DE-independent parametrization of cosmic distances, \citet{Li:2024yrg} also find comparable results, with $\Omega_K=0.09 \pm 0.09$ ($0.06 \pm 0.08$) from \textit{DESI} DR1 BAO data alone (in combination with \textit{PantheonPlus} and CC data): while broadly consistent with our results, we stress that this relies on a specific (albeit quasi-model-independent) three-parameter parametrization, rather than GP reconstruction. The DESI-only GP analysis of~\cite{Liu:2024yib} also finds hints for $\Omega_K>0$, albeit consistent with spatial flatness. Interestingly, in~\cite{Favale:2025mgk} a value of $\Omega_K=-0.143 \pm 0.085$ is inferred from a combination of \textit{DESI} DR2 BAO, \textit{DESY5} SNeIa, and CC data. However, as stressed earlier, their work uses CC to calibrate $M$ and $r_d$, unlike our treatment which is explicitly sensitive only to shape information. This makes a direct comparison of the results difficult, and may contribute to the different constraints.

Other works have conducted null tests for spatial flatness by reconstructing redshift-dependent curvature diagnostics~\citep[see for instance][]{Wu:2022fmr,Dias:2024tpf,Gao:2025ozb,Gong:2025hoy}, with mixed results. In general, since these results cannot be translated into a constraint on $\Omega_K$, they cannot be straightforwardly compared to ours. We also note that the curvature null diagnostic analysis of~\cite{Gong:2025hoy} finds a possible low-redshift deviation which, if confirmed, would appear to go in the direction of an open Universe.

Finally, a qualitatively different comparison is to model-dependent analyses which explicitly assume a given cosmological model, typically non-flat $\Lambda$CDM or extensions thereof, typically in the DE sector. Examples of analyses of this type are those of~\cite{Wu:2024faw,Chen:2025mlf,Chaudhary:2025uzr,Capozziello:2025lor,Chudaykin:2025lww,Yadav:2025wbc,Wang:2025hvo,Wu:2026qog,Giare:2026oti,Chudaykin:2026amr}, several of which report a preference for $\Omega_K>0$, depending on the model and dataset combination adopted. Analyses including CMB data can constrain $\Omega_K$ to the ${\cal O}(10^{-3})$ level, a level of sensitivity which is two orders of magnitude stronger than ours, which however comes at the expense of assuming a specific model. Therefore, these results cannot be directly compared to our late-time, nonparametric constraints on $\Omega_K$. 

Interestingly, a generic finding of these works is how the indications for $\Omega_K>0$ can be substantially weakened (and potentially removed) by introducing additional freedom in the DE sector~\citep{Wu:2024faw,Wu:2026qog,Giare:2026oti}. This is precisely what makes our analysis especially timely, since our methodology is explicitly constructed so as not to rely on a specific DE parametrization, insofar as the late-time expansion history is sufficiently smooth. The persistence of the hints for $\Omega_K>0$ across all three baseline combinations suggest that the post-\textit{DESI} indications for an open Universe are unlikely to be solely an artefact of an assumed, overly restrictive late-time expansion history. In light of the elephant in the room, i.e.\ the Hubble tension, it is important to stress once more that our results are independent of the absolute SNeIa, BAO, and CC calibration.

Another important aspect of our results is that $\Omega_m$, $\sum m_{\nu}$, and $\tau$ do not enter our analysis: $\tau$ is completely irrelevant to our probes, while any effect of $\Omega_m$ and $\sum m_{\nu}$ is implicitly captured by our nonparametric reconstruction. Therefore, the degeneracies affecting these parameters, which complicate the interpretation of model-dependent analyses, play no role in our results. As a note of caution, this does \textit{not} necessarily imply that our results are unrelated to the ``CMB-BAO tension'', which in $\Lambda$CDM manifests as a tension in $\Omega_m$~\citep{Sakr:2023hrl,Baryakhtar:2024rky,Pedrotti:2024kpn,Colgain:2024mtg,Weiner:2026sfm,Shlivko:2026jxa,Schoneberg:2026buf}, since the same redshift-dependent features or systematics present in BAO and SNeIa data could affect both our analysis and model-dependent ones. What we can say for sure is that, if such a relation exists, it certainly does not arise from an assumed value or prior on $\Omega_m$, which does not enter our analysis.

For our analysis, we use the distances and covariance matrices provided by the individual SN~Ia compilations and the BAO data from DESI. There are potential sources of redshift-dependent systematic uncertainty, such as dust properties and progenitor evolution with redshift~\citep{Dhawan:2024gqy}, that are difficult to quantify given current SNeIa data. It will therefore be important to test how future work constraining $\Omega_K$ is impacted by the different sources of systematics. Finally, we further remind the reader that nonparametric is not synonymous with assumption-free. Our analysis still makes a number of assumptions, including but not limited to homogeneity and isotropy on sufficiently large scales, a constant spatial curvature parameter, and a sufficiently smooth expansion history, itself a necessary condition for us to carry out a GP reconstruction. Moreover, systematics which affect the redshift dependence of BAO, SNeIa, and/or CC measurements will inevitably affect our analysis (as well as model-dependent ones). On the other hand, an overall redshift-independent rescaling~\citep[see for instance][]{Pedrotti:2025ccw} would have no effect on our results, which are only sensitive to the shape of the expansion history and not its amplitude.

\section{Conclusions}
\label{sec:conclusions}

We have presented an updated nonparametric determination of spatial curvature using exclusively late-time geometrical probes. Across our three SNeIa compilations, our baseline combinations with \textit{DESI} DR2 BAO data consistently show a mild preference for an open Universe, reaching up to the $\sim 3\sigma$ level, while remaining broadly robust to the choice of expansion rate probe and redshift coverage. Crucially, these constraints are independent of the absolute calibration of the expansion rate and distance scale, which is particularly valuable in the context of the Hubble tension, as disagreements on the absolute distance scale do not propagate into our inference. Our results therefore provide complementary support for recent hints of $\Omega_K>0$, independent of CMB data or the assumption of a parametric dark energy model, warranting future investigations and applications of our method as new late-time geometrical data arrives.

\vspace{-0.5cm}
\section*{Acknowledgements}
H.C.T.\ and S.D.\ are supported by UK Research and Innovation (UKRI) under the UK government’s Horizon Europe funding Guarantee EP/Z000475/1.  S.V.\ acknowledges support from the Istituto Nazionale di Fisica Nucleare (INFN) through the Commissione Scientifica Nazionale 4 (CSN4) Iniziativa Specifica ``Quantum Fields in Gravity, Cosmology and Black Holes'' (FLAG). This publication is based upon work from the COST Action CA21136 ``Addressing observational tensions in cosmology with systematics and fundamental physics'' (CosmoVerse), supported by COST (European Cooperation in Science and Technology).
\vspace{-0.5cm}

\section*{Data Availability}
The data underlying this article will be shared upon reasonable request to the corresponding author (H.C.T.). The analysis code and associated products underlying the work will be made publicly available following publication. In any case, they will be shared upon reasonable request to the corresponding author.

\bibliographystyle{mnras}
\bibliography{omegak} 

\label{lastpage}
\end{document}